\documentclass[preprint, aps, nofootinbib]{revtex4}
\usepackage{CJK}
\usepackage{graphicx}
\usepackage{amsmath}
\usepackage{amsfonts}
\usepackage{amssymb}
\usepackage{color}%
\usepackage{dcolumn}
\usepackage{slashed}
\usepackage{multirow}
\providecommand{\U}[1]{\protect\rule{.1in}{.1in}}
\newcommand{\f}{\begin{equation}}
\newcommand{\ff}{\end{equation}}
\newcommand{\fa}{\begin{eqnarray}}
\newcommand{\ffa}{\end{eqnarray}}

\begin{document}
\title{Lifshitz holographic superconductor in Ho\u{r}ava-Lifshitz gravity}
\author{Cheng-Jian Luo $^{1,2}$}
\email{rocengeng@hotmail.com}
\author{Xiao-Mei Kuang$^{3}$}
\email{xmeikuang@gmail.com}
\author{Fu-Wen Shu$^{1,2}$}
\email{shufuwen@ncu.edu.cn}
\affiliation{$^{1}$Department of Physics, Nanchang University, Nanchang, 330031, China\\
$^{2}$Center for Relativistic Astrophysics and High Energy Physics, Nanchang University, Nanchang 330031, China\\
$^{3}$ Instituto de F\'isica, Pontificia Universidad Cat\'olica de Valpara\'iso, Casilla 4059, Valpara\'iso, Chile}

\begin{abstract}
We study the holographic phase transition of superconductor dual to a Lifshitz black brane probed by an anisotropic scalar field in the probe limit in Ho\u{r}ava-Lifshitz gravity. With the use of numerical and analytical method, we investigate how the critical temperature of the condensation is affected by the Lifshitz exponent $z$,  $\alpha-$correction term in the action as well as the dimensions of the gravity. We also numerically explore the condensation of the dual operator and optical conductivity of the holographic system. Various interesting properties of the holographic condensation affected by the parameters of model are discussed.
\end{abstract}
\maketitle

\section{Introduction}\label{sec1}
The gauge/gravity duality\cite{maldacena1,agmoo,gkp,witten} that relates strongly interacting gauge theories to theories of gravity
in higher dimensions has attracted increasing attention in recent years. It has opened a new window to study
many different strongly interacting systems due to its remarkable feature that it relates a strong coupling QFT
to a weak coupling gravitational theory. For this reason this duality has received considerable
interest in modelling strongly coupled physics, in particular with a view to possible applications to condensed matter physics(see \cite{hartnoll,herzog,mcgreevy} for reviews).

One of the unsolved mysteries in modern condensed matter physics is the mechanism of the high $T_c$ superconductors. In a series studies\cite{gubser1,gubser2},
Gubser proposed a model where the Abelian Higgs model is coupled to
gravity with a negative cosmological constant. In this model there are solutions that spontaneously
break the Abelian gauge symmetry via a charged complex scalar condensate
near the horizon of the black hole. This model exhibits the key properties
of superconductivity: a phase transition at a critical temperature, where a spontaneous
symmetry breaking of a U(1) gauge symmetry in the bulk gravitational
theory corresponds to a broken global U(1) symmetry on the boundary, and the
formation of a charged condensate. Based on this observation, a holographic superconductor model was proposed
by considering a neutral black hole with a charged scalar and Maxwell sector that do not back react on the geometry\cite{Hartnoll:2008vx,Hartnoll:2008kx}. Since then this correspondence
has received considerable considerations in order to understand several crucial properties of
these holographic superconductors (see Ref. \cite{cai1} for reviews).

On the other hand, anisotropic scaling plays a fundamental role in quantum phase transitions in condensed
matter \cite{cardy,sachdev}. Indeed, many behaviors of the condensed matter systems are
governed by Lifshitz-like fixed points. These fixed points are characterized by the
anisotropic scaling symmetry
\f\label{anisotropic}
t\rightarrow \lambda^z t,\ \ \ x^i\rightarrow \lambda x^i,
\ff
where $z$ is called the dynamical critical exponent and it describes the degree of
anisotropy between space and time. A natural generalization of the gauge/gravity duality is the holographic description of QFTs without the conformal invariance. Such attempts were initiated in\cite{son}, where, motivated by fermions at unitarity, holographic
duals to Galilean conformal field theories with Schr\"{o}dinger symmetry was considered. Latter similar theories with Lifshitz symmetry were proposed in \cite{klm} where it was shown that nonrelativistic
QFTs that describe multicritical points in certain magnetic materials and liquid crystals
may be dual to certain nonrelativistic gravitational theories in the Lifshitz space-time
background\footnote{Generalization to theories with hyperscaling violation is possible\cite{kiritsis,xdong}. For a recent progress, please see Refs. \cite{panzhang,lmx,gv,kym,kofinas,dss,elp,fengg,fl,kpww1,lsachdev,kpww2,br,pdsr} for an incomplete list.}, that is
\f\label{Lifmetric}
ds^2=-r^{2z}dt^2+\frac{dr^2}{r^2}+r^2dx^2_i,\ (i=1,\ldots d).
\ff

Recently Ho\u{r}ava suggested that anisotropic scaling even plays a significant role in quantum gravity itself. Starting with the anisotropic scaling \eqref{anisotropic}, he proposed a power-counting renormalizable theory of gravity, the so-called Ho\u{r}ava-Lifshitz (HL) theory \cite{horava1,horava2}.
 The theory is based on the perspective that Lorentz symmetry should appear as an emergent symmetry at long distances, but can be fundamentally absent at short ones \cite{TP,CN}. In the UV, the system exhibits a strong anisotropic scaling between space and time
with $z\geqslant D$ \footnote{One should be careful that $z$ here should not be confused with that one in the metric\eqref{Lifmetric}.}, while at the IR, high-order curvature corrections become negligible, and the Lorentz invariance (with $z = 1$) is expected to be restored.

Since the HL gravity is itself anisotropic between space and time, it is natural to expect that the HL gravity provides a minimal holographic dual for nonrelativistic
Lifshitz-type field theories. Recent progress confirmed this speculation. In \cite{horava3}  it was shown that the Lifshitz spacetime \eqref{Lifmetric} is a vacuum solution of
the HL gravity in (2+1) dimensions and that the full structure of the $z = 2$ anisotropic Weyl anomaly can be reproduced in dual field theories, while its minimal relativistic
gravity counterpart yields only one of two independent central charges in the anomaly. Our recent works \cite{sw1,sw2} found further evidence to support the
above speculation. In particular, we found all the static (diagonal) solutions of the HL gravity in any dimensions and showed that they give rise to very rich space-time
structures: the corresponding space-times include BTZ-like black holes, the Lifshitz space-times, or Lifshitz solitons.

In this paper, we shall propose a holographic model for superconductors in Lifshitz background in HL gravity. The holographic superconductor in the HL gravity has been explored in \cite{Lin:2014bya}. However, the holographic superconductor in Lifshitz black hole in the HL gravity has not been considered before.
Since the model exhibits anisotropic scaling not only in black hole background, but also in the gravity theory itself, one may expect that there should have some different behaviors in this model. Indeed, our results show that
comparing to the Schwarzschild case\cite{Lin:2014bya}, Lifshitz exponent will hinder the condensation of the second operator to occur, namely, it corresponds to lower critical temperature. While the rule is irregular for the first operator, because the condensation of the first operator tends to divergence as one lower the temperature, which is not the typicality in real superconductor.  Related to the second operator, the energy gap, described by the vanishing real part of conductivity, is also harder to open in the field theory dual to the Lifshitz black hole.  Meanwhile, we will also show that the additional $\alpha$-coupling term makes the scalar hair easier to form, which is the same as that found in Schwarzschild case.

The organization of this paper is as follows: in section \ref{sec2}, we construct our model and give a brief introduction of the background geometry. In section \ref{sec3}, we give a detailed calculations on the condensations and the critical temperature. In order to check our numerical results, we also perform analytic calculations. In section \ref{sec4}, the optical conductivity is calculated numerically. Some interesting results obtained there.
In the last section, we present our main conclusions.

\section{Holographic setup and equations of motion}\label{sec2}
In order to show the non-relativistic property in Ho\u{r}ava-Lifshitz gravity, it is more convenient to employ the Arnowitt-Deser-Misner metric formalism
\begin{equation}\label{eq-metric1}
ds^{2}=-N^{2}dt^{2}+g_{ij}(dx^{i}-N^{i}dt)(dx^{j}-N^{j}dt).
\end{equation}
where $N$, $N^{i}$ and $g_{ij}$ are the lapse function, the shift vector and the metric of the spacelike hypersurface, respectively.
The non-relativistic matters including complex scalar and electromagnetic fields in Ho\u{r}ava-Lifshitz(HL) gravity was firstly proposed in \cite{Kiritsis:2009rx,Kimpton:2013zb}. Later, with the aim to study the holographic superconductor dual to this gravity, the Lagrangian of matters with the lowest order is generalized in \cite{Lin:2014bya} as
\begin{eqnarray}\label{eq-action}
S&=&\int\,dt\,dx^{d+1}N\sqrt{g}\left(\frac{1}{4}{\cal L}_1+2{\cal L}_2\right),\\ \nonumber
{\cal L}_1&=&\frac{2}{N^{2}}g^{ij}(F_{0i}-F_{ki}N^{k})(F_{0j}-F_{\ell j}N^{\ell})-F_{ij}F^{ij}-\beta_{0}-\beta_{1}a_{i}B^{i}-\beta_{2}B_{i}B^{i},\\ \nonumber
{\cal L}_2&=&\frac{1}{2N^{2}}|\partial_{t}\psi-iqA_{0}\psi-N^{i}(\partial_{i}\psi-iqA_{i}\psi)|^{2}-\left(\frac{1}{2}-\alpha\right)|\partial\psi-iqA_i\psi|^{2}-\frac{1}{2}V\left(|\psi|\right),
\end{eqnarray}
where$a_i$ and $\beta_i$ are coupling constants of the scalar and gauge fields. When $a_i=\beta_{i}=0$, the above action with the metric (\ref{eq-metric1}) will reproduce the minimal coupling action in the pioneer work\cite{Hartnoll:2008vx} with the potential $V(|\psi|)=m^{2}|\psi|^{2}$. In this paper we would like to limit to this case.

It was addressed in \cite{Lu:2009em} that Schwarzschild spacetime is one of the solutions in Ho\u{r}ava-Lifshitz gravity. Recently,
the authors of \cite{Alishahiha:2012iy} proposed that the HL gravity also admits asymptotically Lifshitz solution
\begin{equation}\label{eq-metric2}
ds^2=-r^{2z}\xi(r)dt^{2}+\frac{dr^2}{r^2\xi(r)}+r^2dx^2_i~~\mathrm{with}~~ \xi(r)=1-\left(\frac{r_0}{r}\right)^{d+z}.
\end{equation}
It should be noted that the dynamic exponent $z$ should not be confused with the one inherent in this theory.
The temperature of this black brane is given by $T_{h}=\frac{d+z}{4\pi}r_0^z$.

In Einstein gravity, the Lifshitz dynamic exponent plays an interesting role in the condensation of holographic superconductor\cite{Bu:2012zzb,Lu:2013tza,Su:2014ghso}.
Thus, it is interesting to see the effect in Ho\u{r}ava-Lifshitz gravity and generalize the holographic superconductor study dual to Schwarzschild black brane \cite{Lin:2014bya} into  the Lifshitz brane (\ref{eq-metric2}).

Similarly, we take the ansatz of matters as $A_{\mu}=(\phi(r),0,0,0)$ and $\psi=\psi(r)$. In the probe limit, we directly variate the action and obtain the equations of motion for the matter fields as
\begin{eqnarray}
\label{eq-eomphi}0&=&\phi''+\frac{d+1-z}{r}\phi'-\frac{2\psi^{2}}{r^{2}\xi(r)}\phi,\\
\label{eq-eompsi}0&=&\psi''+\left(\frac{z+d+1}{r}+\frac{\xi'(r)}{\xi(r)}\right)\psi'+\left(\frac{2}{r^{2}\xi(r)}+\frac{\phi^{2}}{r^{2z+2}\xi^{2}(r)(1-2\alpha)}\right)\psi,
\end{eqnarray}
where we have substituted the metric (\ref{eq-metric2}) and set the mass parameter  as $m^{2}=-2+4\alpha$.

Consequently, we can study the boundary behavior of the matter fields. Near the horizon, $r\rightarrow r_0$, the scalar field satisfies the relation $\psi(r_0)=-\frac{d+z}{2}\psi(r_0)'$, while the U(1) gauge field can be set to be $\phi(r_0)=0$ due to the regularity near horizon.

Near the boundary, i.e., $r\rightarrow\infty$, the behavior of  $\phi$ is
\begin{eqnarray}
\phi&=&\mu-\frac{\rho}{r^{d-z}}+\cdot\cdot\cdot ~~\mathrm{for}~~d\neq z,\\\nonumber
\phi&=&\mu-\rho\log{r}+\cdot\cdot\cdot  ~~\mathrm{for}~~d=z.
\end{eqnarray}
According to the gauge/gravity dictionary, $\mu$ and $\rho$ in the above equations are dual to the chemical potential and charge density in the dual field theory. The boundary behavior of  $\psi$ is
\begin{equation}
\psi=\frac{\psi_1}{r^{\Delta_-}}+\frac{\psi_2}{r^{\Delta_+}}
\end{equation}
with $\Delta_{\pm}=\frac{d+z}{2}\pm\frac{1}{2}\sqrt{(d+z)^{2}-8}$.
Then the condensate of the scalar operator $O$ in the boundary field theory, which is
dual to the scalar field, is given by $<O_i>=\sqrt{2}\psi_i$. Both of them may play the role of source and the other denotes the
expectation value of vacuum with dimension $\Delta_+$ or $\Delta_-$. Thus, the features of the boundary field theory can be read off by studying the asymptotic behavior of matter fields.

\section{Critical temperature and condensation}\label{sec3}
In this section, we shall  solve the equations (\ref{eq-eomphi}) and (\ref{eq-eompsi}) to extract the properties of the phase transition. We will focus on the critical temperature $T_c$ where the scalar hair comes up and the condensation $<O_i>$, i.e., the strength of the dual scalar operator.  Specially, we mainly study the effects of  $\alpha$ correction, Lifshitz dynamical exponent,  and the dimensions of the gravity on $T_c$ and $<O_i>$.

\subsection{Numerical results}
\begin{figure}[h]
  \centering
  \includegraphics[width=.4\textwidth]{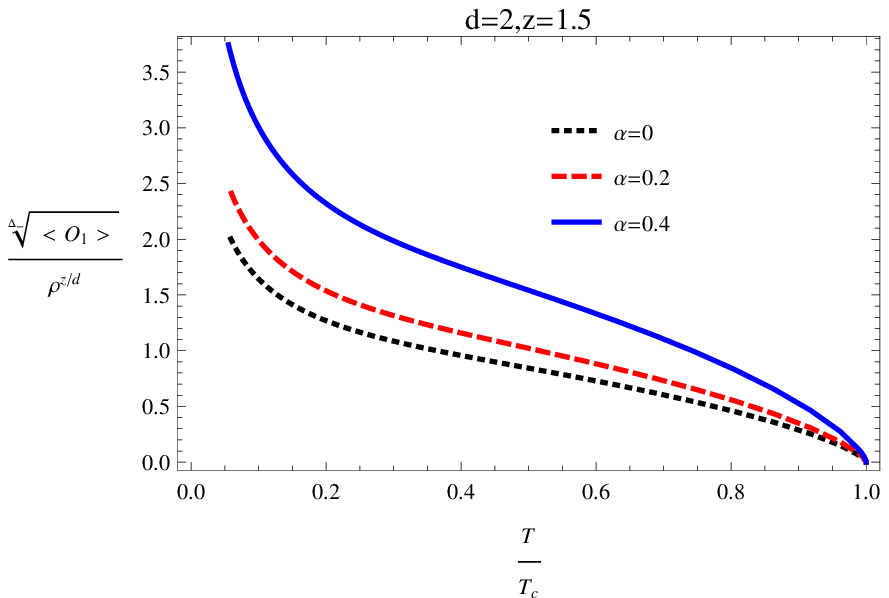}\hspace{1cm}
  \includegraphics[width=.4\textwidth]{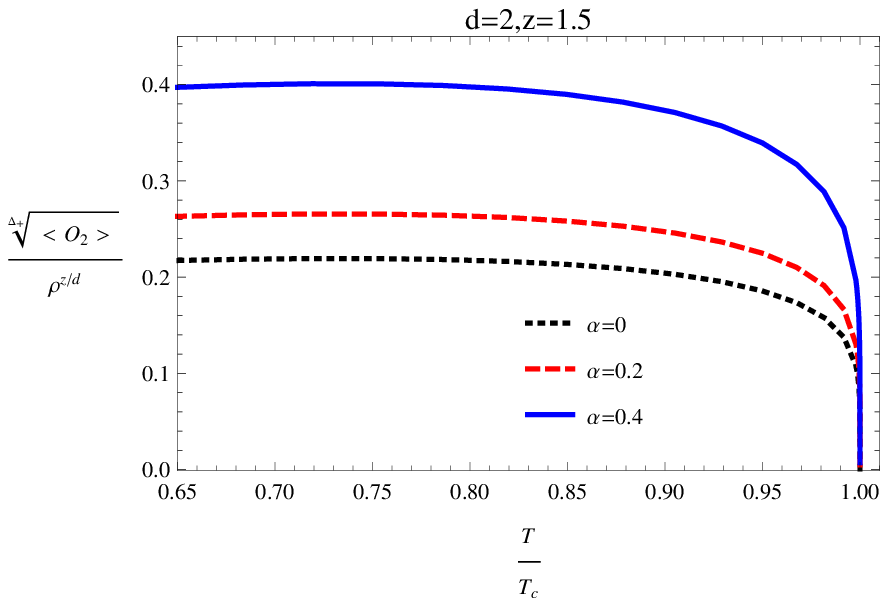}
  \caption{Condensation with $d=2, z=1.5$ for different $\alpha$: dotted line for $\alpha=0$, dashed line for $\alpha=0.2$ and solid line for $\alpha=0.4$.}\label{fig-d2z1p5a}
\end{figure}

\begin{figure}[h]
  \centering
  \includegraphics[width=.4\textwidth]{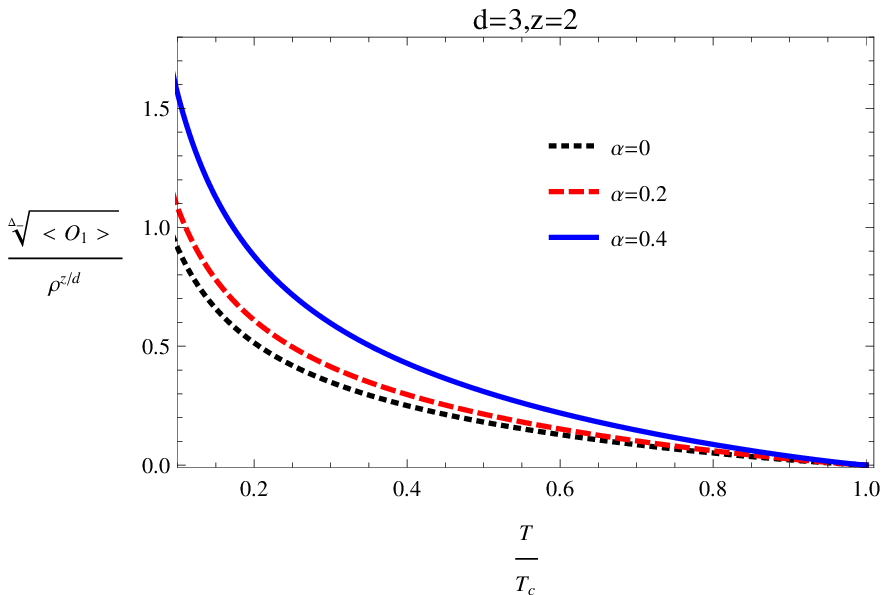}\hspace{1cm}
  \includegraphics[width=.4\textwidth]{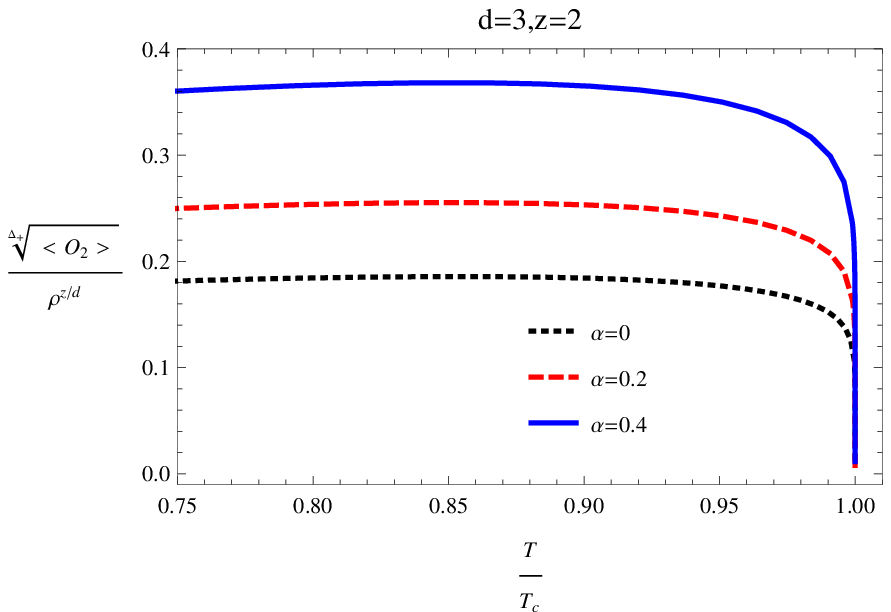}
  \caption{\label{fig-d3z2a}Condensation with $d=3, z=2$ for different $\alpha$: dotted line for $\alpha=0$, dashed line for $\alpha=0.2$ and solid line for $\alpha=0.4$.}
\end{figure}

Firstly, we use shooting method to numerically solve the equations of motion from the horizon to boundary. As the temperature decreases below a critical value $T_c$, the scalar can survive, meaning a phase transition from the normal black hole to a hairy black hole. The strength of condensation becomes large as we continue lowering the temperature. The numerical results are summarized as follows. The critical temperatures for the operators affected by $\alpha$ in asymptotical Lifshitz geometry are shown in TABLE \ref{table-d2z1p5a} and TABLE \ref{table-d3z2a}. We see that in both dimensions $d=2$ and $d=3$,
 larger $\alpha$ corresponds to higher critical temperature, namely, it makes the condensation easier to occur. The strength of condensation related to the parameters in the tables can be found in  FIG. \ref{fig-d2z1p5a} and FIG. \ref{fig-d3z2a}. The positive
 $\alpha$ term correction brings in stronger condensation. These phenomenons are consistent with that observed in AdS background in Ho\u{r}ava-Lifshitz gravity\cite{Lin:2014bya}.
\begin{table}\renewcommand{\arraystretch}{1}\addtolength{\tabcolsep}{10pt}
  \centering
  \begin{tabular}{|c|c|c|c|c|}
   \hline
    $\alpha$& \multicolumn{2}{|c|}{$ <O_{1}>$}& \multicolumn{2}{|c|}{$<O_{2}>$}\\ \hline
    0 & $0.2247(N)$&$0.2247(A)$ & $0.0365(N)$&$0.0359(A)$\\ \hline
    0.2 & $0.2721(N)$&$0.2722(A)$& $0.0442(N)$ &$0.0435(A)$\\ \hline
    0.4 & $0.4109(N)$ &$0.4109(A)$& $0.0667(N)$&$0.0657(A)$ \\  \hline
  \end{tabular}
\caption{\label{table-d2z1p5a}Critical temperature $T_{c}/\rho^{z/d}$ for $<O_{1}>$ and $<O_{2}>$ respectively with different $\alpha$ for fixed $d=2$ and $z=1.5$. `N' means that the result is numerical while `A' denotes that it is analytical.}
\end{table}
\begin{table}\renewcommand{\arraystretch}{1}\addtolength{\tabcolsep}{10pt}
  \centering
  \begin{tabular}{|c|c|c|c|c|}
    \hline
    $\alpha$& \multicolumn{2}{|c|}{$ <O_{1}>$}& \multicolumn{2}{|c|}{$<O_{2}>$}\\ \hline
    0 & $1.6432(N)$ &$1.6415(A)$& $0.0384(N)$ &$0.0556(A)$\\ \hline
    0.2 & $1.9482(N)$ &$1.9463(A)$& $0.0668(N)$ &$0.0660(A)$\\ \hline
    0.4 & $2.8098(N)$ &$2.8070(A)$& $0.0963(N)$ &$0.0951(A)$\\ \hline
  \end{tabular}
  \caption{\label{table-d3z2a}Critical temperature $T_{c}/\rho^{z/d}$ for $<O_{1}>$ and $<O_{2}>$ respectively with different $\alpha$ for fixed $d=3$ and $z=2$.`N' means that the result is numerical while `A' denotes that it is analytical.}
\end{table}
\begin{table}\renewcommand{\arraystretch}{1}\addtolength{\tabcolsep}{10pt}
  \centering
  \begin{tabular}{|c|c|c|c|c|}
    \hline
    $z$& \multicolumn{2}{|c|}{$ <O_{1}>$}&\multicolumn{2}{|c|}{$ <O_{2}>$}\\ \hline
    1 & $0.2255(N)$&$0.2249(A)$ & $0.1184(N)$&$0.1170(A)$\\ \hline
    1.5 & $0.2246(N)$&$0.2247(A) $& $0.0365(N)$&$0.0359(A)$\\ \hline
    2 & $0.6645(N)$& $0.6645(A)$& $0.0289(N)$&$0.0284(A)$\\ \hline
  \end{tabular}
 \caption{\label{table-d2a0z}Critical temperature $T_{c}/\rho^{z/d}$ for $<O_{1}>$ and $<O_{2}>$ respectively with $d=2$ and $\alpha=0$ for different z. `N' means that the result is numerical while `A' denotes that it is analytical.}
\end{table}
\begin{table}\renewcommand{\arraystretch}{1}\addtolength{\tabcolsep}{10pt}
  \centering
  \begin{tabular}{|c|c|c|c|c|}
    \hline
    $z$& \multicolumn{2}{|c|}{$ <O_{1}>$}&\multicolumn{2}{|c|}{$ <O_{2}>$}\\ \hline
    1 & $1.5693(N)$&$1.5690(A)$ & $0.2012(N)$&$0.1987(A)$ \\ \hline
    1.5 & $2.1579(N)$&$2.1563(A)$ & $0.1315(N)$&$0.1299(A)$ \\ \hline
    2 & $1.9482(N)$&$1.9463(A)$ & $0.0668(N)$&$0.0660(A)$ \\ \hline
  \end{tabular}
\caption{\label{table-d3a0p2z}Critical temperature $T_{c}/\rho^{z/d}$ for $<O_{1}>$ and $<O_{2}>$ respectively with $d=3$ and $\alpha=0.2$ for different z.`N' means that the result is numerical while `A' denotes that it is analytical.}
\end{table}
\begin{figure}[h]
  \centering
  \includegraphics[width=.4\textwidth]{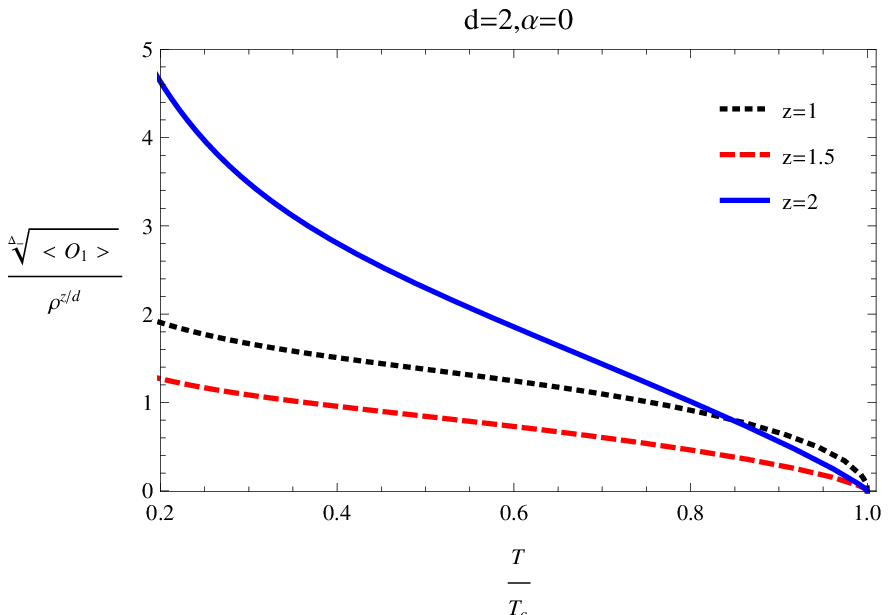}\hspace{1cm}
  \includegraphics[width=.4\textwidth]{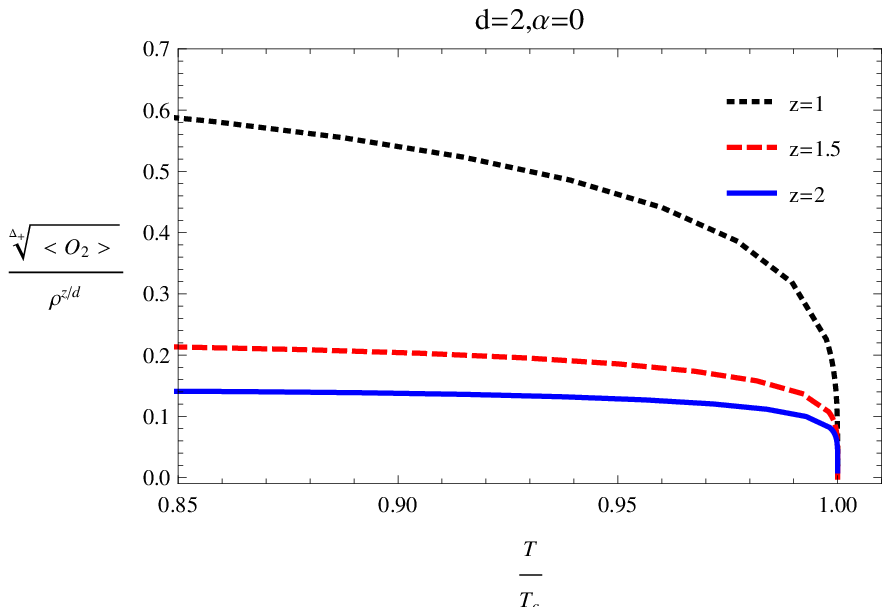}
  \caption{Condensation with $d=2$ and $\alpha=0$: Dotted line ($z=1$),  Dashed line ($z=1.5$), Solid line ($z=2$).}
  \label{fig-d2a0z}
\end{figure}
\begin{figure}[h]
  \centering
  \includegraphics[width=.4\textwidth]{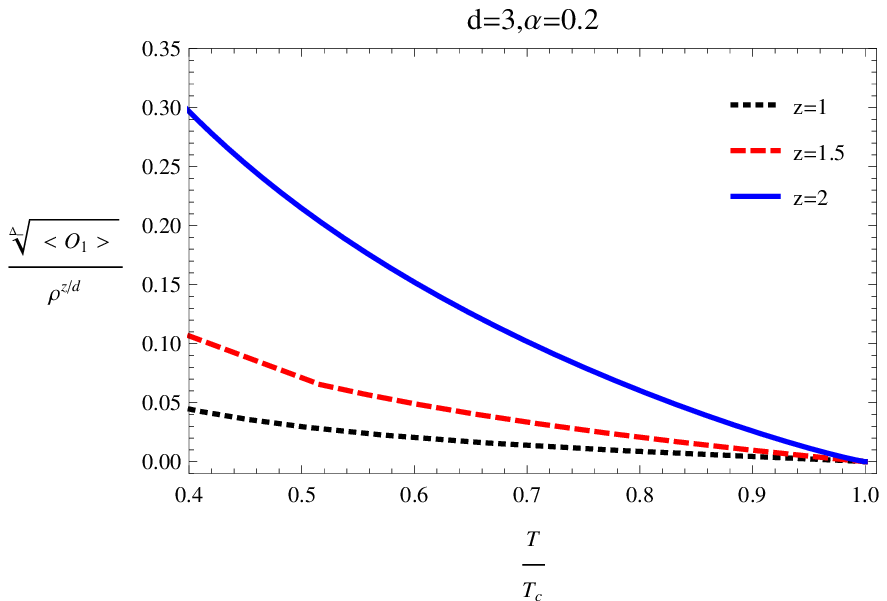}\hspace{1cm}
  \includegraphics[width=.4\textwidth]{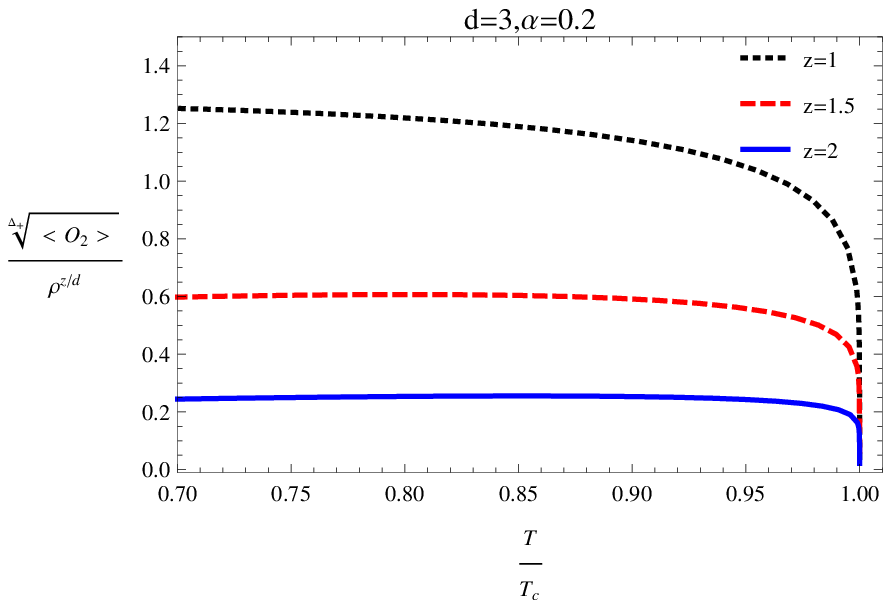}
  \caption{In this figure we can find three lines with $d=3, \alpha=0.2$: Dotted line ($z=1$), Dashed line ($z=1.5$), Solid line ($z=2$).}\label{fig-d3a0p2z}
\end{figure}

We move on to study the effects of the Lifshitz dynamical exponent $z$ on the phase transition. The critical temperatures with different $z$ for fixed dimension and $\alpha$ are listed in TABLE \ref{table-d2a0z} and TABLE \ref{table-d3a0p2z}. In both cases, the critical temperature for $<O_2>$ becomes lower as $z$ increases, meaning that larger $z$ makes the condensation difficult to form. Accordingly, the condensation $<O_2>$ for larger $z$ is weaker as shown in FIG. \ref{fig-d2a0z} and FIG. \ref{fig-d3a0p2z}.
The related results are consistent with the effect of Lifshitz case observed in \cite{Lu:2013tza}. Namely, for the second operator, the condensation is always harder to occur in the field theory dual to Lifshitz black hole than that dual to Schwarzschild black hole in HL gravity. However, for the operator $<O_1>$, the effect of $z$ on the critical temperature and the condensation are irregular. The reason is not clear, but it may attribute to that the first operator does not have a good dual because the condensation is tending to be divergent when the temperature is lowered,  as it discussed in \cite{Hartnoll:2008kx}.

\subsection{Analytical results}
In this subsection we would like to further confirm our numerical critical temperature with the use of Sturm-Liouville method which was first proposed to calculate $T_c$ in holographic in \cite{Siopsis:2010uq}. It is worthy noting that this proposal works well in many cases\cite{Li:2011xja,Jing:2011vz,Gangopadhyay:2012am,Kuang:2013oqa}.

For simpleness, we now redefine the coordinate  $y=r_0/r$, then the above field equations \eqref{eq-eomphi} and \eqref{eq-eompsi} become
\begin{equation}
\psi''(y)+\frac{z+d-1+y^{d+z}}{y(-1+y^{d+z})}\psi'(y)+\left(\frac{2}{y^{2}(1-y^{d+z})}+\frac{y^{2z}\phi^{2}}{y^{2}r_{0}^{2z}(-1+y^{d+z})^{2}(1-2\alpha)}\right)\psi(y)=0,
\end{equation}
\begin{equation}
\phi''(y)+\frac{1+z-d}{y}\phi'(y)+\frac{2\psi^{2}}{y^{2}(-1+y^{d+z})}\phi(y)=0,
\end{equation}
where $0<y\leqslant1$, and $y=1$ is the location of the horizon while $y=0$ denotes the boundary.

Near the boundary ($y\rightarrow0$), we have the asymptotic behaviors for the fields as
\begin{equation}
\psi\approx\psi^{(\pm)}y^{\Delta_{\pm}}
\end{equation}
while
\begin{equation}
\phi\approx\mu-\rho y^{d-z} ~\mathrm{for}~d-z>0~\mathrm{and}~ \phi\approx\mu+\rho\log{y}~\mathrm{for}~d-z=0.
\end{equation}
Note that at the horizon $y=1$, we have the behavior $\psi(1)=-\frac{d+z}{2}\psi(1)'$ and $\phi(1)=0$ as we analyzed in Section \ref{sec2}.

At the critical temperature $T_{c}$, $\psi=0$, then the Maxwell equation becomes
\begin{equation}
\phi''(y)+\frac{1+z-d}{y}\phi'(y)=0,
\end{equation}
whose solution is
\begin{equation}
\phi(y)=\lambda r_0^{z}(1-y^{d-z}) ~\mathrm{for}~d-z>0,
\end{equation}
\begin{equation}
\phi(y)=\lambda r_0^{d}\log{y} ~\mathrm{for}~d-z=0
\end{equation}
where $\lambda$ is defined as  $\lambda=\rho/ r_{0}^{d}$.

Sequently, as the system approach the critical point, i.e., $T\rightarrow T_{c}$, the field equation for the scalar field $\psi$ becomes
\begin{align}
&\psi''+\left(\frac{1-d-z}{y}-\frac{(d+z)y^{d+z-1}}{(1-y^{d+z})}\right)\psi'+\left(
\frac{y^{2z-2}\lambda^{2}(1-y^{d-z})^{2}}{(1-2\alpha)(1-y^{d+z})^{2}}+\frac{2}{y^{2}(1-y^{d+z})}\right)\psi=0£¬
\end{align}
for $d-z>0$, while, for $d-z=0$, it becomes
\begin{align}
&\psi''+\left(\frac{1-d-z}{y}-\frac{(d+z)y^{d+z-1}}{(1-y^{d+z})}\right)\psi'+\left(
\frac{y^{2z-2}\lambda^{2}(\log{y})^{2}}{(1-2\alpha)(1-y^{d+z})^{2}}+\frac{2}{y^{2}(1-y^{d+z})}\right)\psi=0.
\end{align}

At the boundary, we define
\begin{equation}
\psi(y)=\frac{<O>}{\sqrt{2}r_{0}^{\Delta}}y^{\Delta}F(y),
\end{equation}
where $F(y)$ is normalized to $F(0)=1$ and the boundary condition $F'(0)=0$.
Hereafter, for $d-z>0$, the  equation of motion for the scalar is
\begin{align}
&F''+\left(\frac{1+2\Delta-d-z}{y}+\frac{(d+z)y^{d+z-1}}{-1+y^{d+z}}\right)F'+\left(\frac{\Delta^2y^{d+z-2}}{-1+y^{d+z}}\right)F\nonumber
\\&=-\frac{y^{2z-2}\lambda^{2}(1-y^{d-z})^{2}}{(1-2\alpha)(-1+y^{d+z})^{2}}F,
\end{align}
and for $d-z=0$, it is
\begin{align}
&F''+\left(\frac{1+2\Delta-d-z}{y}+\frac{(d+z)y^{d+z-1}}{-1+y^{d+z}}\right)F'+\left(\frac{\Delta^2y^{d+z-2}}{-1+y^{d+z}}\right)F\nonumber
\\&=-\frac{y^{2z-2}\lambda^{2}(\log{y})^{2}}{(1-2\alpha)(-1+y^{d+z})^{2}}F.
\end{align}
The above two equations for the function $F$ can be converted into
\begin{equation}
(T(y)F')'-[P(y)-\lambda^2Q(y)]F=0,
\end{equation}
where
\begin{align}
&T(y)=y^{2\Delta+1-d-z}(1-y^{d+z}),\nonumber
\\&P(y)=\Delta^2y^{2\Delta-1}
\\&Q(y)=y^{2\Delta-1-d+z}\frac{(1-y^{d-z})^{2}}{(1-y^{d+z})(1-2\alpha)} ~\mathrm{for}~d-z>0,
\\ \mathrm{or}~&Q(y)=y^{2\Delta-1-d+z}\frac{(\log{y})^{2}}{(1-y^{d+z})(1-2\alpha)} ~\mathrm{for}~d-z=0.
\end{align}

With the use of Sturm-Liouville theory, the eigenvalue of $\lambda$ can be found as the minimum value of
\begin{align}\label{eq-lambda}
\lambda^{2}=\frac{\int_{0}^{1}dy[T(y)F'^{2}(y)+P(y)F^{2}(y)]}{\int_{0}^{1}Q(y)F^{2}(y)dy}.
\end{align}
In order to get the explicit integral result of the above formula, we choose
\begin{align}\label{eq-F}
F(y)=1-\beta y^{2}.
\end{align}
Then, substituting (\ref{eq-F}) into (\ref{eq-lambda}), we can integrate out $\lambda^{2}$.  Choosing proper $\beta$, we can minimize the expression and get $\lambda_{min}$, thus,  the critical temperature is
\begin{align}\label{eq-Tclambda}
T_{c}=\frac{d+z}{4\pi}r_{0}^z=\frac{d+z}{4\pi\lambda_{min}^{z/d}}\rho^{z/d}.
\end{align}

For example, when $d=2,\alpha=0$ and $z=1$, we have the integration (\ref{eq-lambda}) for the first operator $<O_{1}>$
\begin{align}
\lambda^2=\frac{6-6\beta+10\beta^2}{2\sqrt{3}\pi-6\ln3+4(\sqrt{3}\pi+3\ln3-9)\beta+(12\ln3-13)\beta^2}.
\end{align}
$\beta\approx0.239$ minimizes the above equation, then  (\ref{eq-Tclambda}) gives us the critical temperature
$T_c\approx0.225\sqrt{\rho}$ which is exact the value in AdS case. The analytical critical temperatures for the two operators in all discussed cases with different $\alpha$ and $z$ are also summarized in the corresponding tables.  After comparing the results in all the tables, it is obvious that the analytical results and  the numerical values match very well.

\section{Optical conductivity}\label{sec4}
To calculate the optical conductivity, we should consider the perturbed Maxwell field
$A_{i}=\delta_{i}^{x}e^{-i\omega\,t}A_{x}(r)$, then $A_x(r)$ satisfies the following equation
\begin{equation}
A_{x}''+\left(\frac{z+d-1}{r}+\frac{\xi'(r)}{\xi(r)}\right)A_{x}'+\left(\frac{\omega^{2}}{r^{2z+2}\xi^{2}(r)}-\frac{2(1-2\alpha)\psi^{2}}{r^{2}\xi(r)}\right)A_{x}=0.
\end{equation}
Near the horizon, we impose the ingoing condition $A_x\propto \xi(r)^{\frac{-i\omega}{d+z}}$.
Near the boundary, the general solution to the fluctuation is
\begin{equation}
A_{x}=A_{x}^{(0)}+A_{x}^{(1)}y^{d+z-2}+\cdots.
\end{equation}
Note that in some cases, such as $d=z$  or $d=3$ and $z=1$, etc., there is  a logarithmic term proportional to $\ln {(Cr)}$
with $C$ a constant. For more details, please refer to the references \cite{Lu:2013tza,Kuang:2015mlf}.
Subsequently, with the suitable renormalization, the optical conductivity of the dual holographic superconductor, can be calculated as\cite{Hartnoll:2008vx,Kuang:2015mlf}
\begin{equation}
\sigma(\omega)=\frac{d+3z-4}{i\omega}\frac{A_{x}^{(1)}}{A_{x}^{(0)}}~~\mathrm{or}~~\sigma(\omega)=\frac{d+3z-4}{i\omega}\frac{A_{x}^{(1)}}{A_{x}^{(0)}}+\frac{i\omega}{d+3z-4}.
\end{equation}

\begin{figure}
  \centering
  \includegraphics[width=.4\textwidth]{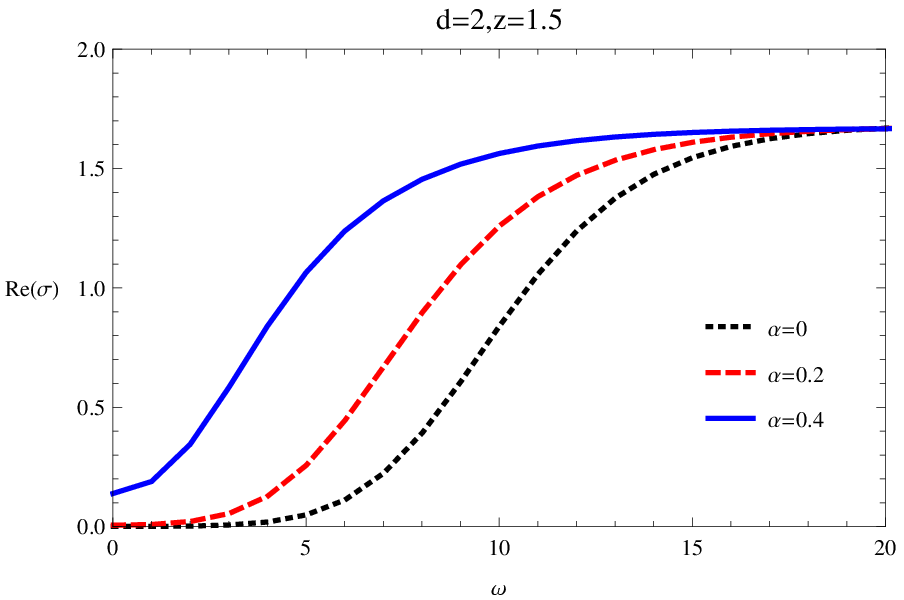}\hspace{1cm}
  \includegraphics[width=.4\textwidth]{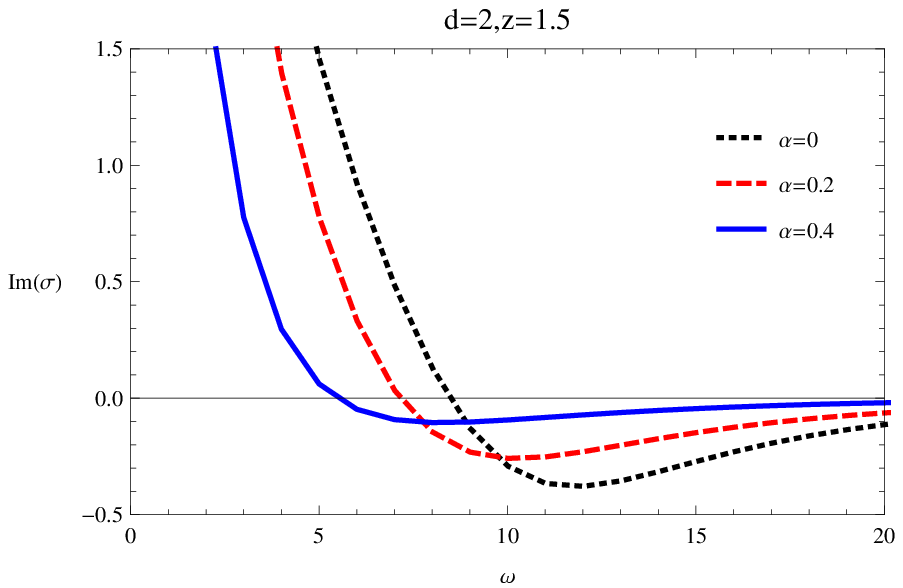}
  \caption{The conductivity at $T/T_c=0.25$ with $d=2$ and  $z=1.5$ for $\alpha=0, 0.2$ and $0.4$.}\label{fig-conduc1}
\end{figure}

\begin{figure}
  \centering
  \includegraphics[width=.4\textwidth]{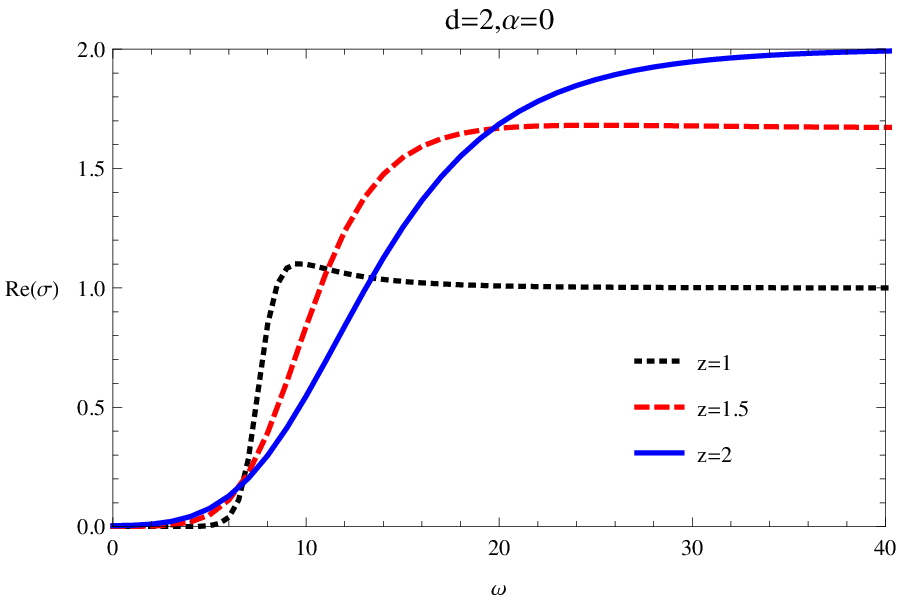}\hspace{1cm}
  \includegraphics[width=.4\textwidth]{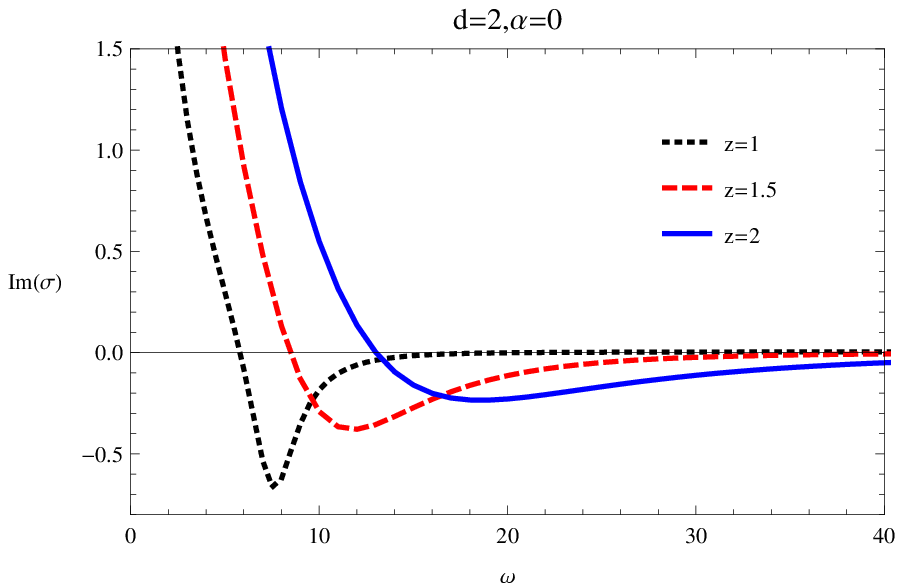}
  \caption{The conductivity at $T/T_c=0.25$ with $d=2$ and  $\alpha=0$ for $z=1, 1.5$ and $2$.}\label{fig-conduc2}
\end{figure}

In FIG. \ref{fig-conduc1} and FIG.  \ref{fig-conduc2}, we show the real part and imaginary part of optical conductivity  for
$T/T_{c}\approx0.25$ in $d=2$ geometry\footnote{Results for $d=3$ are similar.}. More concretely, FIG. \ref{fig-conduc1} shows that with fixed $z=1.5$, larger $\alpha$ makes the gap in the real part of $Re(\sigma)$ narrower,
which is consistent with the result for $z=1$ as found in \cite{Lin:2014bya}.  However, since the spacetime is Lifshitz instead of AdS, the large frequency is not scaled to be unit as it always occurs in Lifshitz case. With fixed $\alpha=0$, as the dynamical exponent $z$ increases, the low frequency behavior of the real part of $\sigma$ is enhanced, namely, more shift from zero, while the high frequency behavior is also improved, which agrees well with the observations in \cite{Sun:2013zga}.


\section{Conclusion and Discussion}\label{sec5}

We have investigated the holographic superconductor dual to a Lifshitz black hole in Ho\u{r}ava-Lifshitz gravity.  For generality, an $\alpha$ correction term was introduced into the action of the matter field. We mainly focused on studying the effect of the $\alpha$ coupling and the Lifshitz exponent on the condensation and the optical conductivity of the dual superconductor. Our main results can be summarized as follows.  Firstly, in Lifshitz case, the $\alpha$ coupling increases the critical temperature so as to promote the hairy scalar to form for both operators, and it decreases the energy gap described by the real part of the conductivity at lower temperature. These observations are consistent with that found in Schwarzschild case in HL gravity. Then, comparing to the dual AdS case,  Lifshitz exponent always suppresses the second operator to condensate, while the rule is irregular for the first operator because it is not the typical superconductor due to the divergence at zero temperature limit.  Related to the second operator, the energy gap is smaller in the field theory dual to Lifshitz black hole because the asymptotical behavior is modified.

In this work, we ignored the backreaction of matter fields into the gravity and study in the probe limit. It is interesting to take the backreaction into account and check the properties obtained in the present paper. Without loss of generality, generalization of this work to the fermionic case is possible. It is expected that fermionic sector may exhibit much more fruitful structures  in this context,
especially the phase transition disclosed in \cite{fangk}.  Moreover, it is also interesting to study the formation of the crystalline geometries\cite{shu1} in the present context, to see which configuration of the lattice tends to form.

\begin{acknowledgments}
This work was supported in part by the National Natural Science Foundation of China under Grant No. 11465012(FWS)), the Natural Science Foundation of Jiangxi Province under Grant No. 20142BAB202007(FWS) and the 555 talent project of Jiangxi Province(FWS). The work is also partly supported by FONDECYT grant No.3150006(XMK).
\end{acknowledgments}

\end{document}